\documentclass[preprint,aps,graphicx]{revtex4}
\usepackage{graphicx}
\usepackage{dcolumn}
\usepackage{bm}

\begin{document}
\title{Stability of fermionic Feshbach molecules in a Bose-Fermi
mixture}
\author{Alexander V.Avdeenkov$^{(1)}$ Daniele C.
E. Bortolotti$^{(2,3)}$, and John.L.Bohn$^2$}
\address{(1)Institute of Physics and Power Engineering
Obninsk,Kaluga region,RUSSIA, 249033\\
(2)JILA, NIST, and Department of Physics,~University of Colorado,
Boulder, CO 80309-0440\\ (3)LENS and Dipartimento di Fisica,
Universit\'a di Firenze, Sesto Fiorentino, Italy}


\begin{abstract}
In the wake of successful experiments in Fermi condensates,
experimental attention is broadening to study resonant
interactions in  degenerate Bose-Fermi mixtures.  Here we consider the
properties and stability of the fermionic molecules that can be
created in such a mixture near a Feshbach resonance (FR).  To do
this, we consider the two-body scattering matrix in the many-body
environment, and assess its complex poles.  The stability
properties of these molecules strongly depend on their
centre-of-mass motion, because they must satisfy Fermi statistics.
At low centre-of-mass
momenta the molecules are more stable than in the absence of the
environment (due to Pauli-blocking effects), while at high
centre-of-mass momenta nontrivial many body effects render them somewhat
less stable.
\end{abstract}
\maketitle

\section{introduction}

Nowadays the physics of cold atomic gases  and their mixtures provide
extraordinary opportunities both to test theoretical models and to
predict new phenomena. The most remarkable effects arise when
a resonant interaction can be engineered between pairs of atoms,
by means of a magnetic-field Feshbach resonance (FR).  The effects of
resonant interactions are by now widely studied both theoretically and
experimentally for both Bose and  Fermi systems. The resonant
interaction for  two-component Fermi systems enabled the
realization of BCS-BEC crossover~\cite{jin1,jin2,ketterle,grimm}.
The Bose system with a resonant interaction revealed such phenomena as
the ``Bose-nova''~\cite{wieman}.

To combine particles of both symmetries, and to consider a resonant
interaction in a Bose-Fermi (BF) mixture, is a logical next step
in this field.  Recently, experiments have observed
interspecies FR's between bosonic atoms and fermionic
atoms~\cite{jin,modugno,simoni,ferliano}.
The BF FR brings new possibilities to the physics of quantum gases
such as boson-mediated Cooper
pairing~\cite{kagan,matera,bijlsma}, phase separation
\cite{molmer,amoruso}
and the creation of a superfluid of fermionic polar molecules~\cite{baranov}.

For trapped BF mixtures with a tunable interspecies interaction,
one can expect a variety of regimes that are defined both by
the interaction strength and by the densities of the two species.
For example, the BF mixture may generate a new kind of a crossover upon
crossing the resonance region. On the atomic
side of the resonance, BCS  superfluidity may be induced by the
exchange of a boson density fluctuation, whereas superfluidity
will likely disappear on the molecular side, as the molecules fill
a Fermi sea.  Further, if the crossing of the resonance
can be accompanied by a process that produces ground state fermonic
molecules (say, by a stimulated Raman process) then the resulting molecules
would possess strong dipolar interactions that may restore superfluidity.

But beyond this, the behaviour of the atom pairs in the resonant
BF mixture depends subtly on the momentum of the pairs.  Roughly
speaking, for a single, free pair of atoms, a true molecular bound
state exists on one side of the resonance, denoted as the ``negative
detuning'' side.  On the other, ``positive detuning,'' side, the
pair is not rigorously bound but may exhibit resonant scattering.
For free molecules this demarcation at zero detuning between bound
and resonant states is clearly independent of the pair's centre-of-mass
momentum.  In the many-body environment, however, this situation
changes dramatically.  We will show that slowly moving molecules
can be stable against decay even on the positive detuning side of
resonance.  The reverse is also true: pairs that are moving fast enough
will become unstable and exhibit only resonances on the negative
detuning side, even though their two-body analog would be
completely stable. This unusual behaviour is connected to the fact
that the pairs are themselves fermions, and must obey the correct
Fermi statistics.  This is of course different from the case of
either a boson-boson or fermion-fermion resonance, where the
pairs are always bosons.

Both a quantitative and
qualitative explanation of the resulting phenomena requires
thorough implementation the two-body and many-body physics.
Fortunately, a mean-field approximation is expected to give
reasonable physical insight of physics near a FR
~\cite{perali,avdeenkov,diener}.
In this article we will consider only  one aspect of BF
mixture near FR, namely how the stability of a composite fermionic
molecule  will be affected by the many-body medium.  To
do this we will assess the poles of the many-body T-matrix
of a molecular pair propagating in the many-body medium.


\section{Poles of the T-Matrix}

\subsection{Two-body case}
The model we will use in the following is somewhat well studied,
having been introduced in similar
contexts \cite{holland_ren,stoof2}; in Ref.~\cite{me} it was applied to the
system at hand. We will therefore abstain from providing
many of the details, and lean substantially on the results obtained in
those works.

The FR in two-body scattering is identified by parameterizing
the dependence of the scattering length on the detuning, $\nu$,
which represents the molecular binding energy in the case of an
infinitely narrow resonance, and, in practice, is tuned using magnetic
fields.  We will then use the  following
parameterization for the scattering length
\begin{equation}
\label{scattering_length}
a = a_{bg}-{ m_{bf} g^2 \over 2 \pi \nu},
\end{equation}
where $a_{bg}$ is the background scattering length away from resonance,
$m_{bf}$ is the boson-fermion reduced mass, and $g$ is a parameter
representing the width of the resonance.
For concreteness, we consider parameters appropriate  FR
that has been observed in in  K-Rb \cite{jin,ferliano}.
Using the parameters defined in Ref. \cite{me}, we have
$a_{bg}=-202$ $a_0$, $\delta_B=5.1 \times 10^{-5}$ K/G,
and $\Delta_B=1$G, $B_0=513$G, whereby the working  parameters become
$\nu=\delta_B (B - B_0)$ and $g=\sqrt{V_{bg} \delta_B \Delta_B}$,
where $ V_{bg}={2 \pi a_{bg} \over m_{bf}}$ .
In terms of these parameters, the low-energy two-body scattering
$T$-matrix near a resonance takes the familiar form \cite{landau}
\begin{eqnarray}
\label{2body}
 T_{2B}(E)=-\frac{2\pi\hbar} {m_{bf}}
\frac{1}{-a^{-1}+r_0m_{bf}E-i\sqrt{2 m_{bf}E}},
\end{eqnarray}
where $r_0=-2\pi/m_{bf}^{2}g^{2}$ is the effective range of the
interaction resulting from the underlying model.

\begin{figure}
\centerline{\includegraphics[width=0.75\linewidth,height=0.6\linewidth,
angle=-0]{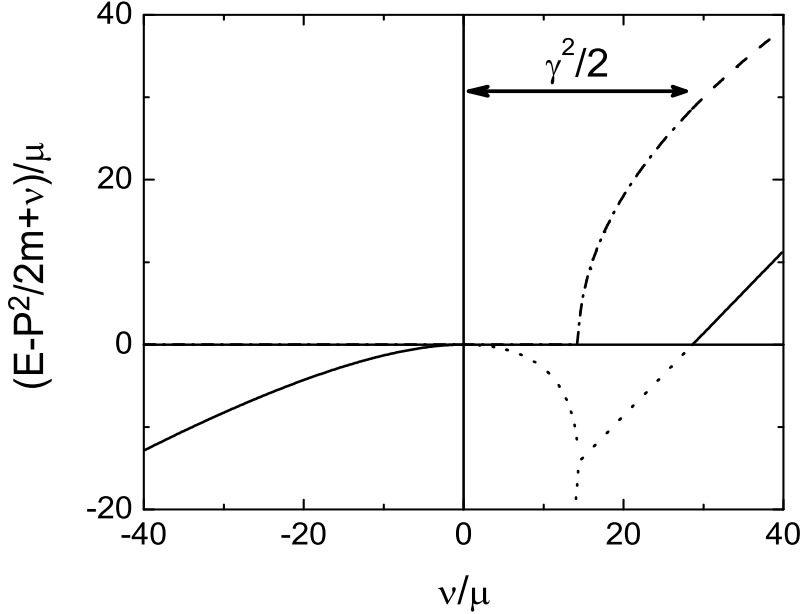}}
\caption{Complex poles of the two-body T-matrix, as a function of
detuning.  Solid and dashed lines denote the real and imaginary
parts of physically relevant poles, respectively. The dotted and
dash-dotted lines are real and imaginary parts of unphysical poles,
respectively. }
\label{twobody_poles}
\end{figure}

Bound states and resonances of the two-body system are identified
in the structure of poles of (\ref{2body}) in the complex plane.
This is illustrated in Figure \ref{twobody_poles}, where real
and imaginary parts
of the poles' energies are plotted as a function of detuning $\nu$.
Here $\nu$ is given in units of $\mu$, which is the same as the
chemical potential in the many-body example below.  We normalize
$\nu$ in this way for easier comparison between the two-body
and many-body results.
For $\nu < 0$, the two-body system possess a true bound state,
whose binding energy is denoted by the solid line.  In this case,
the pole of (\ref{2body}) occurs for real energies.  This bound state
vanishes as the detuning goes to zero, which in fact is what defines
the zero of detuning.

For positive detunings $\nu > 0$, there is no longer a true bound
state, but there may be a scattering resonance.  This resonance appears
for detunings $\nu / \mu > g^4 m_{bf}^3 / 4 \pi^2$, about 28 in
Fig.~\ref{twobody_poles} and its position is shown as a solid
line, representing the real part of the pole.  At the same detuning,
there emerges a positive imaginary part of the pole energy (dashed line),
which denotes the energy width of the resonance.  Interestingly, for
detunings $0 < \nu / \mu < g^4 m_{bf}^3 / 4 \pi^2$, the poles of
(\ref{2body}) are purely
imaginary, and the imaginary part is negative.  These poles stand for
physically meaningless solutions to the Schr\"{o}dinger equation,
in which the amplitude in the resonant state would grow exponentially
in time, rather than decay.  These poles do not therefore identify
any particular features in the energy-dependent cross section of the
atoms.
There is a characteristic detuning scale on which these events occur.
This scale is given by the width parameter
$\gamma^2 = g^2M_{bf}^{3/2}/\sqrt{2}\pi$ \cite{andreev}, also
indicated in the figure.

\subsection{Many-body case}

In a many-body environment, the $T$-matrix and its poles depend on the
centre-of-mass momentum. The importance of taking into account
this dependence was demonstrated in~\cite{levin} for the BCS-BEC
crossover at $T \neq 0$ when non-condensed fermion pairs lead to
pseudogap effects above $T_c$ and non BCS- behaviour below.
 The influence of the many- body medium on a
Feshbach resonance between two fermions was demonstrated
in~\cite{bruun} for a composite boson as well.
In the BF mixture the composite object is  a fermion,
so its momentum dependence  cannot be omitted even for $T = 0$.
 Thus the $T$-matrix near the FR in a BF medium must be considered as a
 function of densities of both the Bose and Fermi subsystems as well as the
centre-of-mass motion of a given BF pair. We have found that
T-matrix poles and residues are quite sensitive to all these
ingredients.


The $T$-matrix $\hat{T}$ of the system in the many-body medium is defined
by the Lippmann-Shwinger equation:
\begin{equation}
\hat{T}=\hat{g}\hat{D}\hat{g^{\dag}}+\hat{g}\hat{D}\hat{g^{\dag}}\hat{G_{B}}
\hat{G_{F}} \hat{T}
\label{tmat}
\end{equation}
where $\hat{D}$ is the renormalized molecular Green function, and
$\hat{G_{B/F}}$ is the boson/fermion renormalized Green function, all
of which are defined in terms of their well known \cite{fetter} unrenormalized
counterparts $\hat{D^0}$, and $\hat{G^0}_{B/F}$ by the self-consistent
set of equations
\begin{eqnarray}
\label{general}
&&\nonumber \hat{D}=\hat{D^0}+\hat{D^0}\hat{\Pi}\hat{D}
\\
\nonumber
&&\hat{G_B}=\hat{G_B^0}+\hat{G_B^0}\hat{g}\hat{G_F}\hat{D}\hat{g^{\dag}}\hat{G_B}
\\
\nonumber
&&\hat{G_F}=\hat{G_F^0}+\hat{G_F^0}\hat{g}\hat{G_B}\hat{D}\hat{g^{\dag}}\hat{G_F}
\end{eqnarray}
where $\hat{\Pi}$ is the molecular self energy, and the Green functions
for fermions $\hat{G_F}$, bosons $\hat{G_B}$
and molecules $\hat{D}$ represent $2\otimes2$ functions
containing their anomalous (pairing) parts.  All quantities are
functions of the energy $E$ and centre-of-mass momentum $P$
of the molecules.  The complete solution
to these equations is beyond current computational capabilities,
except perhaps by Monte Carlo methods.

We therefore make a few simplifying assumptions, namely
we account for propagation of the atomic fermions and
bosons using their free Green functions only
(i.e., setting $\hat{G_F} \approx \hat{G_F^0}$ and
$\hat{G_B} \approx \hat{G_B^0}$ ). An important consequence of these
choices, is that the many-body $T$-matrix is approximated by its ladder
series, which means that
$\hat{\Pi}\approx g \hat{G^0_B}\hat{G^0_F}g^{\dag}.$ This standard
approximation has the property of being exact in the 2 body limit,
where Eq.~(\ref{2body}) satisfies (\ref{tmat}). This implies that the
two body physics is accounted for exactly in the many body problem.

Another appealing characteristic of this approach is that the
molecular self energy $\Pi$ can be calculated exactly
\cite{albus}, leading to the following expression
\begin{eqnarray}
&\Pi(E,P)= { g^2 \over \pi^2}  m_{bf}\Lambda - {g^2
    \over 4 \pi^2}  m_{bf} k_f-
{g^2 \over 8 \pi^2} \left(
{m_b k_f^2 \over P} \right. \nonumber \\
& -  \left. {m_{bf}^2 P \over
  m_b} -{m_b D
  \over P} \right)
\ \ {\rm ln}\left({(k_f +P\ m_{bf} /m_{b})^2-D \over (k_f -P \ m_{bf} /m_{b})^2-D
} \right) \nonumber \\
&+{g^2 \over 4 \pi^2} m_{bf}\sqrt{D}\ \ {\rm ln} \left({(k_f+\sqrt{D})^2-(P\ m_{bf}
  /m_b})^2\over {(k_f-\sqrt{D})^2-(P \ m_{bf}  /m_b})^2\right),
\label{selfe}
\end{eqnarray}
where $D=\sqrt{2m_{bf}(E-P^{2}/2(m_{f}+m_{b})+\mu)}$, and
$\Lambda$ is an ultraviolet cutoff, which can be regularized by a
shift in the detuning $\nu \rightarrow \nu-{ g^2 \over \pi^2}
m_{bf}\Lambda$.

Further approximations include considering  a homogeneous system where
 the  density of the fermionic
subsystem is much smaller than that of  the bosonic one.
This is important for treating the bosonic subsystem as
almost unperturbed by the fermionic one.
Finally, we disregard the boson-boson interaction, and
include only the resonant part
of the Bose-Fermi interaction,  setting $a_{bg}=0$,  in
 Eq.~(\ref{scattering_length}), though not in the definition of $g$.

The set of equations we obtain with these approximations is therefore
\begin{eqnarray}
T(E)=g^2 D(E),
\label{tfin1}
\end{eqnarray}
where $D(E)$ is the pair propagator defined by:
\begin{eqnarray}
&&D(E)=D^0(E)+D^0(E)\Pi(E) D(E),\\
\nonumber &&\Pi(E)= g^2 G^0_B(E) G^0_F(E),
\end{eqnarray}
which lead to
\begin{eqnarray}
T(E,P)={g^2 \over E-{P^2 \over 2(m_f +m_b)}-\nu+\mu-\Pi(E,P)}.
\label{tfin}
\end{eqnarray}

In order to understand the stability of a BF molecule, we study
the structures of the poles and residues of the T-matrix in
Eq.~(\ref{tfin}).  To determine numerically the poles, we consider
this equation as a system of two nonlinear equations for real and
imaginary parts of the internal energy,
${E-P^{2}/2(m_{f}+m_{b})+\mu}$. The resulting nonlinear system of
equations always has some unphysical solutions which can be
rejected by  the following analysis: i) the residues for physical
solutions must be less than unity;  ii) for the imaginary
solutions  the relative momentum,
$D=\sqrt{2m_{bf}(E-P^{2}/2(m_{f}+m_{b})+\mu)}$ in
Eq.~(\ref{selfe}), should dwell on the lower half of the complex
momentum plane (the second sheet for a complex energy), as for
resonant scattering \cite{newton};
 iii) the sum rule (\ref{sr}, see below) should be fulfilled including
both the discrete and continuum parts.

\begin{figure}
\centerline{\includegraphics[width=0.75\linewidth,height=0.6\linewidth,
angle=-0]{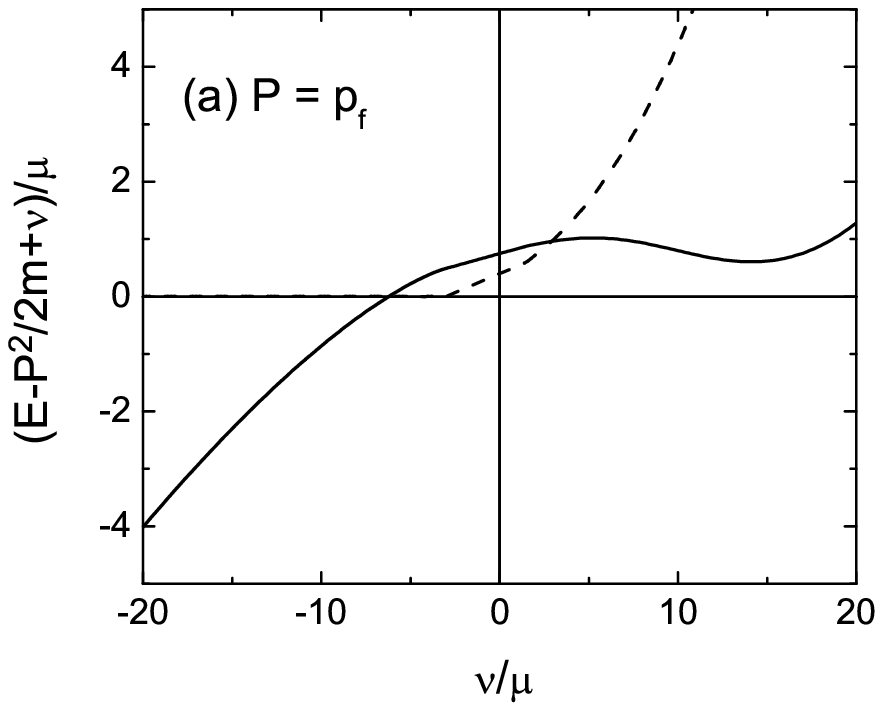}}
\centerline{\includegraphics[width=0.75\linewidth,height=0.6\linewidth,
angle=-0]{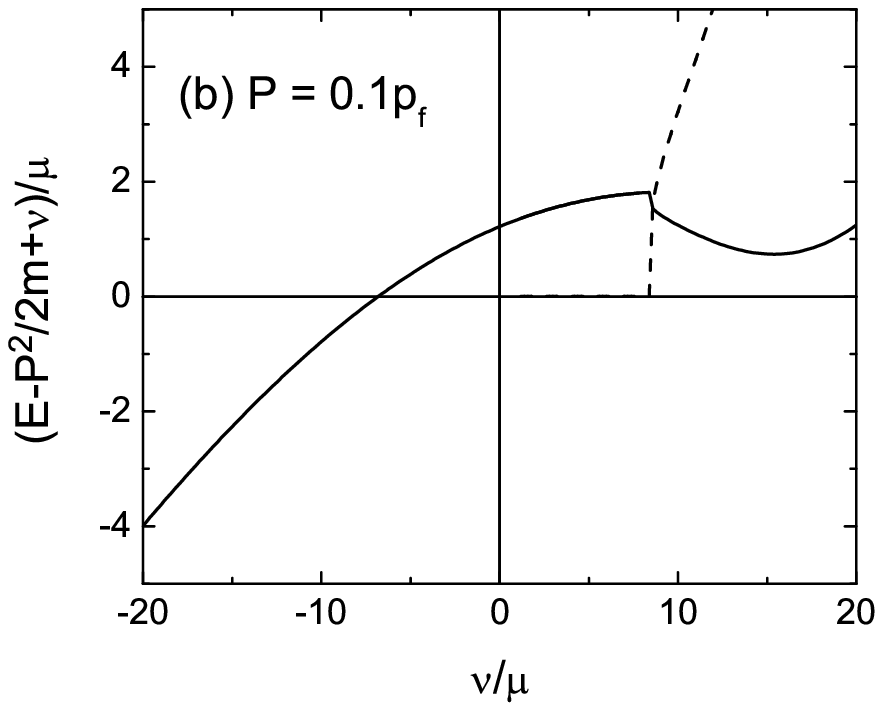}}
\caption{Complex poles of the many-body T-matrix, as a function of
detuning.  Solid and dashed lines denote the real and imaginary
parts of physically relevant poles.  In (a) the centre-of-mass
momentum of the molecule is $P = k_f$, i.e., equal to the Fermi
momentum of the atomic gas.  In (b), this momentum is $P = 0.1 k_f$.}
\label{manybody_poles}
\end{figure}

As an example, we have calculated the poles of (\ref{tfin}) for a
dual species gas with fermion density $10^{13}$ cm$^{-3}$ and
boson density $10^{14}$ cm$^{-3}$.  Moreover, for simplicity, we
assume that the bosons and fermions have the same mass, $m=40$
amu. The results are shown in Figure \ref{manybody_poles} for two
representative centre-of-mass momenta $P$ of the molecules. (Note
the different scale from Figure \ref{twobody_poles}). Unphysical
poles are not shown in this figure.  These results are cast as a
kind of ``binding energy,'' by subtracting the centre-of-mass
kinetic energy, and adding the chemical potential $\mu$.  They can
therefore be compared directly to the two-body results in Fig.
\ref{twobody_poles}.

The structure of these poles is quite different near resonance,
although we stress that far from resonance ($\nu \gg \gamma^2/2$)
they return to the two-body values. A main difference from the
two-body case is that now there may exist physical poles for
any detuning across the resonance.
The appearance of poles depends, however, on the
centre-of-mass momentum $P$ of two atoms. If $P \gg 2k_f$,
the two-body physics will not be
influenced much and the poles structure will be as it is
in Fig.~\ref{twobody_poles}.  We have verified this numerically,
but have not shown it in this paper.

Figure \ref{manybody_poles}(a) shows the case where the centre-of-mass
momentum of the atom pair is equal to the Fermi momentum
of the atomic gas, $P = k_f$.  The criterion of molecular
stability is set by the vanishing of the imaginary part
of the pole.  Figure \ref{manybody_poles}(a) shows that
the imaginary part remains zero until $\nu / \mu \sim -5$,
after which the pair becomes unstable.  Thus, for detunings
$-5 \le \nu / \mu \le 0$, some of the molecules that would have been
stable become de-stabilized in the many-body environment.
They may still be rather long-lived, however.
In the limit of very large momenta, $P \gg 2k_f$, we find that
the pole structure returns to the two-body value.  This makes
sense, since a very rapidly-moving molecule does not interact
strongly with the gas at all.  In particular, for $P \ge 2k_f$,
we find that a ``gap'' reappears, in which no physical poles
exist for some range of positive detuning.  In the $P \rightarrow
\infty$ limit, this gap returns to its two-body value, $\gamma^2/2$.

Figure  \ref{manybody_poles}(b) shows an alternative case in which
the molecular momentum is much smaller than the atomic Fermi
momentum, $P = 0.1 k_f$.  In this case, the imaginary part of the
pole only differs from zero at positive detunings, $\nu / \mu >
8$. Therefore, quite the opposite to the $P = k_f$ case, here the
molecules that would have been unstable are stabilized by the
presence of the many-body environment.  Roughly, this is due to
Pauli blocking of the fermionic atoms into which the molecules
would dissociate.  If such an atom already occupies the state into
which the molecule would drop its atom, then the process is
forbidden.  We discuss this further below.  For any molecule
with $P \le 2k_f$, there exist physical poles at all
detunings.  Thus positive-detuning molecules are always present
in the BF mixture.

\section{Conditions for molecular stability}

To make a more global picture of whether molecules are stable
or not for a given detuning, we can consider the spectral function
for atom pairs.  In general, when the molecular propagator
possess an imaginary part, this part alludes to the decay rate
of the pair due to interactions with the rest of the gas, in the
same sense that the oscillator strength of an atom alludes to
its decay rate by spontaneous emission.  Following a standard approach
\cite{mahan}, the spectral function is defined as
\begin{eqnarray}
\rho(E,P)=-\frac{1}{\pi}ImD(P,E).
\end{eqnarray}
In the case of a true, bound molecular state, the spectral
function reduces to a delta function at the energy of the state $E_0$:
\begin{equation}
\rho (E,P) = 2 \pi Z(P) \delta(E-E_0),
\end{equation}
where the coefficient is given by the ``spectral weight'' function
\begin{eqnarray}
Z(P)=\frac{1}{1-dRe(\Pi(P,E))/dE|_{E_{0}}}.
\end{eqnarray}
Just as for an oscillator strength, the discrete and continuum
parts of the spectral density must satisfy a sum rule:
\begin{equation}
Z(P)+\int dE\rho(E,P)=1
\label{sr}
\end{equation}
for each momentum $P$.  We have explicitly verified the sum rule in
each case we computed, as a test of the numerical procedure, and to
distinguish between physical and unphysical poles of the T-matrix.

\begin{figure}
\label{zcont}
\centerline{\includegraphics[width=.75\linewidth,height=0.6\linewidth,
angle=-0]{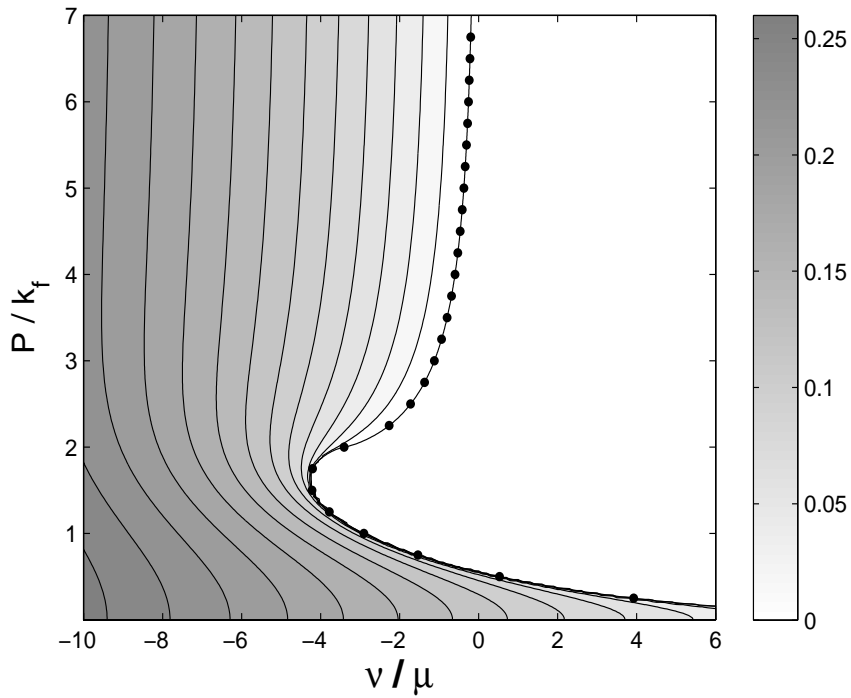}}
\caption{Contours of molecular population under various
combinations of molecular centre-of-mass momentum $P$ and detuning
$\nu$.  The uppermost contour identifies the detuning at which
bound molecules first appear for a given momentum $P$. Numbers
indicate contours with a equal molecule fraction. The dots
represent the result obtained analytically for the critical detuning,
Eq.~(\ref{nucrit}).}
\end{figure}

As mentioned above, the spectral weight $Z$ is associated with a
specific pole $E_0$ of the T-matrix, an energy level
of the system, and represents its
population. Therefore, if $Z$ vanishes, so does the probability of
finding stable molecules in the gas.
 To this end,
Figure \ref{zcont} plots contours of the function $Z(P)$ as a function of
detuning $\nu$ and centre-of-mass momentum $P$ of the atom pairs.
These calculations are performed for fermion and boson densities
 $n_f = 10^{13}$ cm$^{-3}$ and $n_b = 10^{14}$ cm$^{-3}$.
The contour of $Z=0$ thus represents the borderline between
conditions where molecules exist and are stable (to the left of
this line) and where they are unstable to decay (to the right of
this line, in the white region of the graph).

Figure \ref{zcont} thus shows that: i) molecules are still stable
for a continuum of positive detuning when $P$ is small;
ii) molecules that {\it
would have been} stable at negative detuning may not be such
stable at intermediate momenta $P$ (although at small negative
detuning they may possess  small widths); and iii) in the limit $P
\rightarrow \infty$, the borderline between stable and unstable
again returns to zero detuning. Detailed information on the
spectral density can also be used to elicit the momentum
distribution of molecular pairs, a task that will be completed in
future work \cite{Bortolotti}.

Thus far, these are rigorous results, at least within the
simplifying approximations made above.  Once this is done, the
effective dissociation energy of the molecules within the medium
is determined.  The relation between the molecule's total
energy at dissociation and the molecule's
momentum is then easily determined from kinematics, plus
simple considerations on the Pauli blocking of the atomic fermions.
For example, consider the case where the molecule's kinetic energy
is greater than twice the atomic Fermi energy, i.e.,
$P^2/2(m_f+m_b) \ge 2 \times k_f^2/2m$.  At the same time, the molecule is
assumed to be exactly at its dissociation threshold, so that it
could live equally well as a molecule or as two free atoms.
Upon dissociating, each atom would carry away half the energy,
so in particular the fermionic atom is at the top of the Fermi
sea, and this dissociation is not prevented by the Pauli exclusion principle.
The total energy of the molecule at its dissociation threshold
is then determined simply by the molecular kinetic energy, and no
contribution is required from the molecular binding energy.  Thus, if
B represents the internal energy of the pairs relative to threshold,
bound molecules are possible when
\begin{equation}
B \le 0 \;\;\;\;\;
{\rm for} \;\;\;\;\; P \ge \sqrt{{2(m_b+m_f)\over m_f}} k_f.
\label{highP}
\end{equation}

Alternatively, suppose the molecules have less than twice the
atomic Fermi energy, $P^2/2(m_f+m_b) \le 2 \times k_f^2 / 2m$.  Now
it is no longer  guaranteed that the molecules can automatically
decay in the many-body environment, since the fermion's kinetic
energy may lie below the Fermi level of the atomic gas.  In
such a case, the molecule can sustain a {\it positive} internal
energy without dissociating, simply due to Pauli blocking.
To decide how high this binding energy can be, we examine
the conservation of energy and momentum in the dissociation
process:
\begin{eqnarray}
{\bf P} &=& {\bf p}_f + {\bf p}_b  \\
\frac{P^2}{2(2m)} + B &=& \frac{p_b^2}{2m_b} + \frac{p_f^2}{2m_f}.
\label{conservation}
\end{eqnarray}
Here ${\bf p}_f$, ${\bf p}_b$, and ${\bf P}$ are the momenta
of the atomic fermions, atomic bosons, and molecules, respectively.
To ensure that the atomic fermion emerges with the maximum possible
kinetic energy, we consider the case where ${\bf P}$ and ${\bf p}_f$
point in the same direction.  To ensure that $p_f > k_f$, where
$k_f$ is the atomic Fermi momentum, along with (\ref{conservation}),
implies that molecules are stable when
\begin{equation}
B \le {(m_f P-(m_b+m_f)k_f)^2\over 2 m_b\ m_f (m_b+m_f)}
\;\;\;\;\; {\rm for} \;\;\;\;\; P \le \sqrt{{2(m_b+m_f)\over m_f}} k_f.
\label{lowP}
\end{equation}

\begin{figure}
\centerline{\includegraphics[width=.75\linewidth,height=0.6\linewidth,
angle=-0]{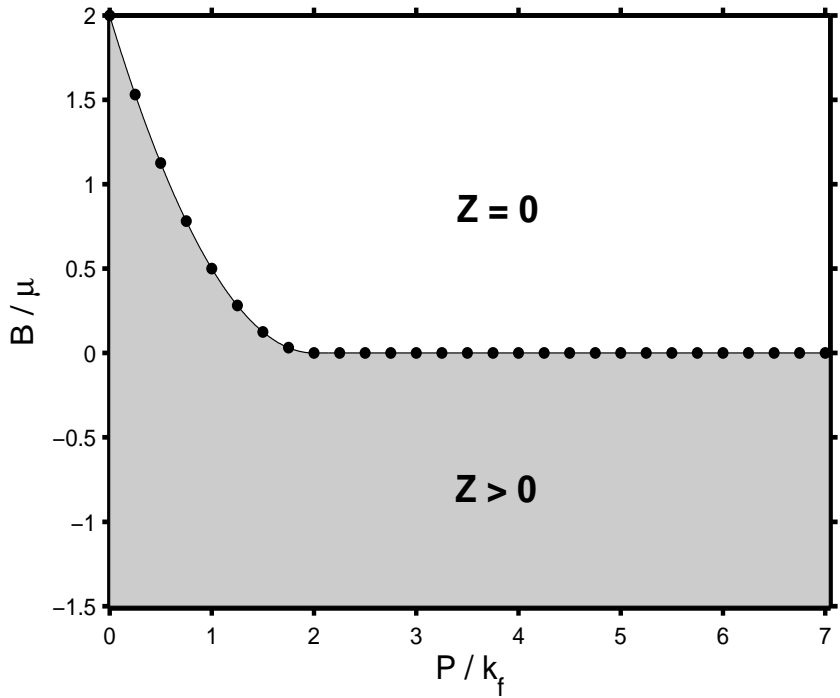}}
\caption{The internal energy that a molecule
  of centre-of-mass momentum $P/k_f$ would require to overcome Pauli
  blocking and dissociate. The shaded area represents numerical
  calculations, and the dots represent the analytical
  kinematic arguments in Eqs.~(\ref{highP},\ref{lowP}).}
\label{bfig}
\end{figure}

Figure \ref{bfig} shows the internal energy
of the molecules evaluated at the stability boundary, as described
above, as a function of centre of mass momentum.
 The solid line in this figure is determined
numerically from the $Z=0$ contour of Fig.~\ref{zcont}.
Subtracting the kinetic energy contribution and chemical potential
from the pole of the T-matrix evaluated on the contour, we obtain the
molecular internal energy. Also shown, as dots,
are the kinematic estimates (\ref{highP},\ref{lowP}).

Analytical expressions for the detuning at
the boundary as a function of centre of mass momentum, plotted as dots
in Fig.~\ref{zcont}, are readily
obtainable analytically, with similar accuracy by inspecting the
denominator of Eq.~(\ref{tfin}). The
total energy of the molecule, measured from the chemical potential,
is, in fact,
given by the pole of (\ref{tfin}). In general this energy can be written
as $E_{0}={P^2 \over 2(m_f +m_b)}-\mu+B$, where B is a complicated
function of all the parameters. However, since $E_{tot}$ is a pole of
(\ref{tfin}), then  $E_{0}={P^2 \over 2(m_f
  +m_b)}-\mu+\nu+\Pi(E_{0},P)$, so
\begin{equation}
\label{nucrit}
\nu_{crit}=B-\Pi \left({P^2 \over 2(m_f
  +m_b)}-\mu+B,P \right).
\end{equation}
Plugging the stability boundary value of B from
Eqs.~(\ref{highP},\ref{lowP}) into
this formula leads to an analytic, albeit complicated,
expression for the critical detuning as a function of centre of mass momentum.

\section{Conclusions}
The constitution and stability of a composite fermion in the
many-body environment is an important part of the BF crossover
regime. Within our approach we have addressed properties of strongly
correlated BF pairs and defined the stability region. We concluded
that in the many-body environment, low momentum molecules can be
stabilized at positive detuning, and intermediate-momentum
molecules can be de-stabilized, existing for  shorter times at
small negative detunings. We also concluded that there is always
a probability to observe molecules at positive detunings,
provided their momenta are less than $\approx 2p_f$, even though
these molecules would not exist as two-body objects.  The way
in which these molecules are distributed will form the basis of
future work.

\acknowledgements This work was supported by the DOE, by a grant
from the W. M. Keck Foundation. A.Avdeenkov acknowledge very
useful discussion with S.Krewald.

\end{document}